# Efficient use of paired spectrum bands through TDD small cell deployments

A. Agustin, S. Lagen, J. Vidal, O. Muñoz, A. Pascual-Iserte, G. Zhiheng*,W.Ronghui*

Signal Processing and Communications Group (SPCOM), Universitat Politècnica de Catalunya, Barcelona, Spain

(*) Huawei Technologies Co., Ltd, Peking, P.R. China

{adrian.agustin, sandra.lagen, olga.munoz, antonio.pascual, josep.vidal}@upc.edu

{guozhiheng, wenronghui}@huawei.com

**Abstract**. Traditionally, wireless cellular systems have been designed to operate in Frequency Division Duplexing (FDD) paired bands that allocates the same amount of spectrum for both downlink (DL) and uplink (UL) communication. Such design is very convenient under symmetric DL/UL traffic conditions, as it used to be the case when the voice transmission was the predominant service. However, with the overwhelming advent of data services, bringing along large asymmetries between DL and UL, the conventional FDD solution becomes inefficient. In this regard, flexible duplexing concepts aim to derive procedures for improving the spectrum utilization, by adjusting resources to the actual traffic demand. In this work we review these concepts and propose the use of unpaired Time Division Duplexing (TDD) spectrum on the unused resources for small eNBs (SeNB), so that user equipment (UEs) associated to those SeNB could be served either in DL or UL. This proposal alleviates the saturated DL in FDD-based system through user offloading towards the TDD-based system. The flexible duplexing concept is analyzed from three points of view: a) regulation, b) Long Term Evolution (LTE) standardization, and c) technical solutions.

# 1 Introduction

The most salient feature in the evolution of mobile services is the overtaken of data services over voice traffic demand, thus requiring the redefinition of current wireless cellular networks and communication standards. Second and third generation of wireless cellular systems were designed under a symmetric traffic assumption as a result of the predominance of voice traffic, and the common technical solution adopted worldwide was the use of paired bands under frequency division duplexing (FDD). The legacy of this assumption has survived in 4G system, even though a TDD frame definition was also early adopted.

New habits of users produce high asymmetries in data traffic demand, i.e. the amount of data transmitted in the downlink (DL) connection is usually much higher than the amount of data in the uplink transmission (UL) [1]. The most conservative measured DL:UL asymmetry ratio across different macro eNBs (MeNBs) is 4:1 [2], due to video downloading and internet browsing. On the other hand, the uploading of shared contents in social media is generating also the opposite tendency, where ratios of 1:4 have been reported in [3]. Such time/space-varying unbalance of traffic affects negatively the spectral efficiency of FDD-based systems: its inflexibility translate in underuse of one band while the other may be congested. This inefficiency could be reduced if adopting unpaired band technologies based on time division duplexing (TDD), in which the use of radio resources can be flexibly adapted as a function of the traffic demands in DL and UL.

This work explores the spectral efficiency improvement of long term evolution (LTE) FDD-based systems under traffic asymmetries by means of the flexible duplexing concept [4]. The proposed solution assumes a deployment of TDD-based small eNBs (SeNB) operating in the resources that are not used by the FDD MeNB, see Figure 1. The following challenges have to be faced:

- *Coexistence of adjacent FDD/TDD systems*. Because of non-ideal transmit filters, adjacent channel interference (ACI) comes up in systems operating in adjacent bands. It can be managed either by imposing a minimum distance between transmit nodes [5], or defining a set of guard bands and power spectrum masks [6].

- *Impact of different TDD-LTE frame pattern configurations*. Conventionally, in TDD modes all MeNBs transmit simultaneously in DL, while all UEs transmit in UL. This has the objective of limiting the active nodes that generate interference on each case. However, flexible use of TDD MeNB/SeNBs entails the decision of its own DL-UL frame pattern, and introduces new types of interference in cellular systems, i.e. MeNBs/SeNBs can be interfered by other MeNBs/SeNBs. If this interference is properly managed, then significant throughput gains are possible in low-to-medium system loads [5].

- *Shared access of the spectrum*. Interference management is important when eNBs with different maximum transmitting power levels are operating on the same radio resources. However, deploying outdoor SeNBs (with a maximum EIRP of 30 dBm and a height below 12 meters) is a simple solution that guarantees that the generated interference is not so important, allowing a large reuse of the spectrum [7].

The proposed solution is investigated in [8] where TDD cognitive SeNBs are allowed to exploit the FDD-DL spectrum. In that case, the TDD SeNB listens to the FDD UL signals in order to detect if there are active FDD-UEs (or MUEs in the following) in the neighboring area. If this is not the case, the UEs associated to the SeNB (SUEs in the following) are allowed to transmit in the FDD-DL band. Furthermore, an implementation of a TDD system in the unused FDD-UL spectrum is proposed in [9], where the interference between the FDD-UL and TDD systems is avoided thanks to a tight time coordination between FDD and TDD systems.

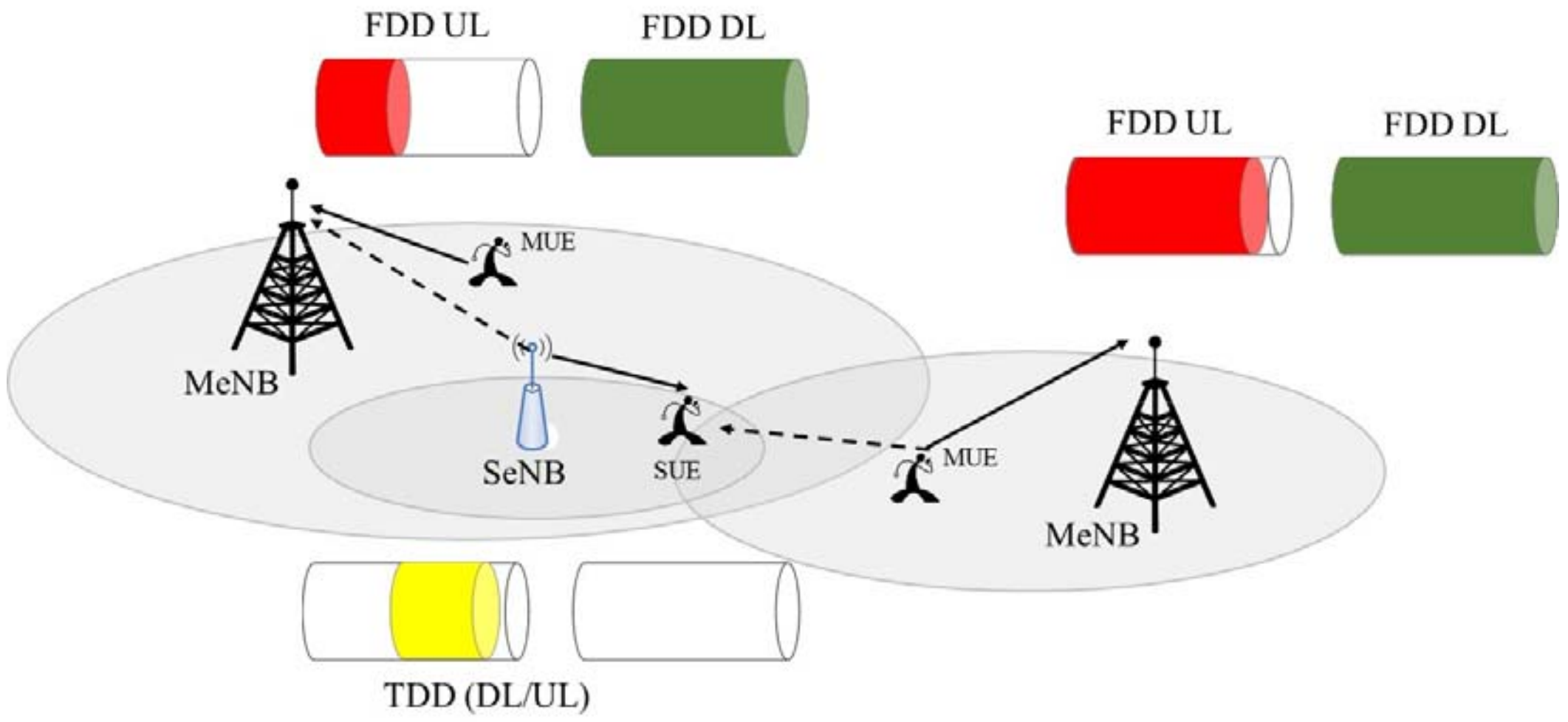

**Figure 1. Example of flexible duplex concept. One macro-cell area with unused resources in the UL, but saturated in the DL. A TDD SeNB is allowed to operate in the unused resources, at the cost of coexisting with the MeNB.**

Through the rest of the paper, we examine alternatives and challenges in the implementation of the flexible duplexing concept in LTE. Specifically, Section 2, details schemes along with the pros and cons of reusing either the FDD-UL or the FDD-DL. Section 3 addresses limitations of the scheme proposed due to the regulation and/or LTE standard constraints. Section 4 reviews technical challenges that come up when deploying multiple TDD SeNBs. Finally, Section 5 underlines some conclusions.

# 2 Flexible use of the paired band

In an FDD-based system a guard band (usually of several MHz) separates the paired UL and DL bands. This constraint imposes a certain spectral gap, that cannot be accommodated in the current underutilized spectrum (usually of few MHz). Fortunately, TDD-based systems are not affected by such limitation so we investigate the deployment of TDD SeNBs operating in the unused spectrum of FDD-based systems. The different options for multiplexing are described in Section 2.1. Section 2.2 discusses how the resource provisioning could be estimated. The benefits of the proposal are shown in Section 2.3.

## 2.1 Flexible duplexing methods in LTE

The options for implementing the flexible duplexing concept depend on whether it is possible to release part of the licensed FDD spectrum, having special consideration to those frequency resources in LTE reserved for control channels. The first option, named efficient in-band use of the licensed spectrum, subsumes the case where the SeNB works in the band of interest, either in orthogonal or non-orthogonal access. The different implementations are depicted in Figure 2 A) opportunistic or non-orthogonal, B) orthogonal in time, C) orthogonal in frequency, and are described in Section 2.1.1. The second approach, addressed in Section 2.1.2 and illustrated in Figure 2 D), assumes that MeNB can make good use of the intra-band component carrier (CC) aggregation LTE concept, adapting the operational bandwidth according to the MeNB traffic demand. MeNB and SeNB work orthogonally in frequency, but in contrast to the approach in Figure 2 C, the FDD-based system can adapt its reserved resources. SeNB can access the channel through frequency multiple access.

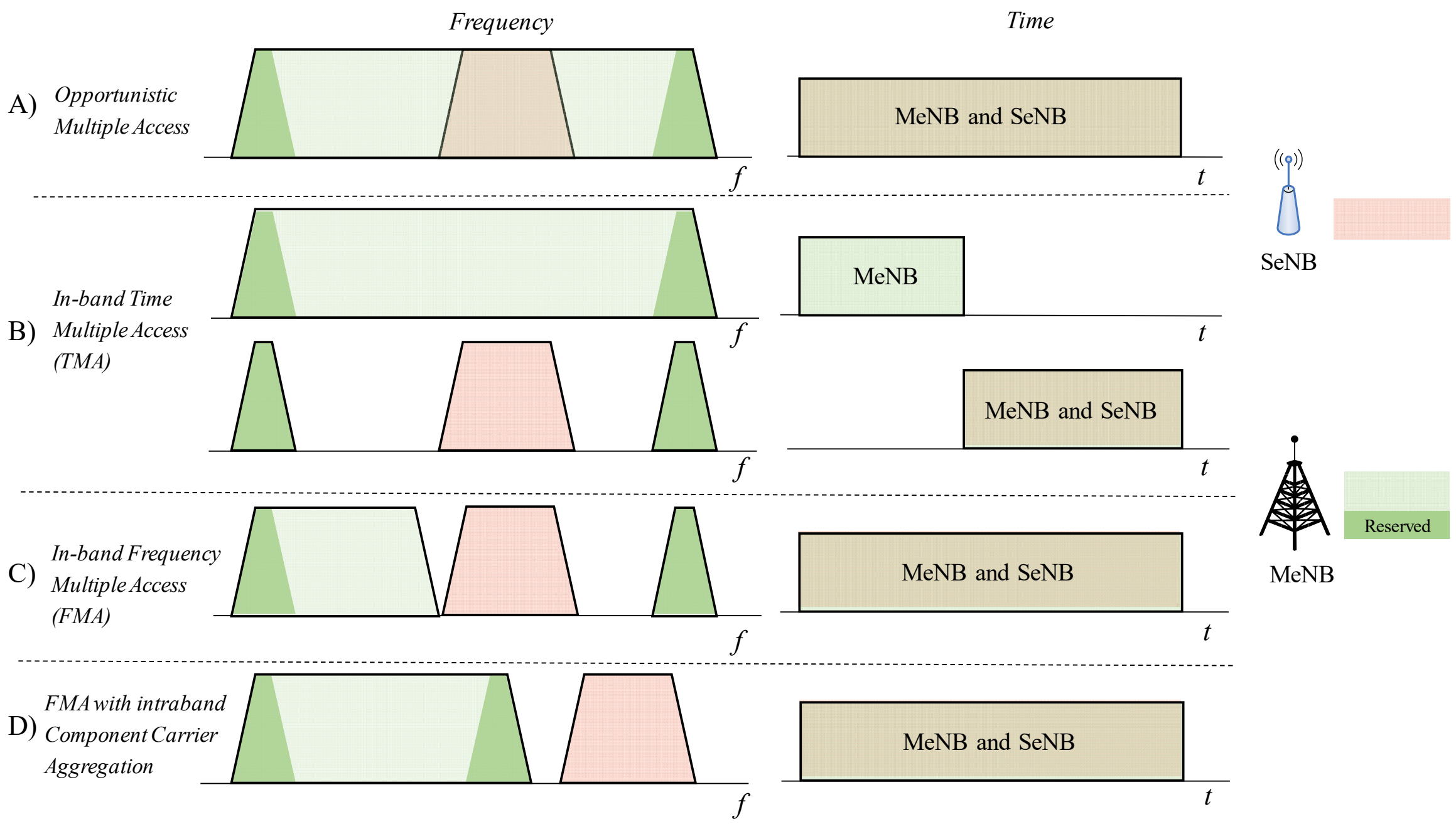

**Figure 2. A), B), C) An overlaid TDD SeNB operates in the unused resources of FDD-UL under opportunistic multiple access (non-orthogonal), time-multiple access and frequency multiple access, respectively. In D), the MeNB adjusts its operational LTE FDD-UL bandwidth using CC aggregation and the TDD SeNB works in the unused resources. Dark-green resources at the left-had side denote reserved frequencies for control-plane communications.**

### 2.1.1 Efficient in-band use of the licensed spectrum

When the SeNB and the MeNB operate in the same band, there are two possible ways of accessing the channel:

*Opportunistic multiple access*. The SeNB is operating in some resources where the MeNB is also operating, see Figure 2-A). The use of outdoor LTE-based SeNBs, working under orthogonal frequency division multiplexing (OFDM), demands an efficient management of intra-subcarrier and inter-subcarrier interference. In the latter case, it comes up when the signals are not properly time-aligned at the receiver side. For example, the signals transmitted in the UL for the TDD SeNB and the ones transmitted from MUEs come from different positions, so if they are not time-corrected in order to be aligned within the cyclic prefix at the MeNB and at the SeNB, the orthogonality between subcarriers is lost.

*Orthogonal multiple access*. The SeNBs might use the MeNB band in time-multiplexing (TMA) or frequency-multiplexing (FMA), see Figure 2-B and C). The second approach requires that the receiver at the MeNB be equipped with analog filters to avoid the undesired sidelobe signal transmission coming from SeNBs in those resources, because it receive filter is configured for the whole operational bandwidth. Therefore, this option comes at the cost of flexibility since unused time frequency resources of the FDD band have to be identified beforehand and be fixed over time.

Reusing the FDD-UL band with the in-band TMA approach demands to pay special attention to physical UL control channels (PUCCHs) allocated at the edge of the band, see dark-green resources in Figure 2. Those resources are used by MUEs to transmit acknowledgements at a predetermined delay following the DL transmission. Therefore, SeNB and MeNB should agree on accessing at those resources at different time instances, or SeNB and MeNB should be equipped with additional analog filters to preserve the adequate reception of the PUCCH at the MeNB. Similarly, the in-band FMA needs that transmitting and receiving nodes are equipped with analog filters, and the in-band interference becomes ACI.

Reusing the FDD-DL band is more complicated due to the current frame structure. The reserved frequency resources for control-plane communications are found in the central part of the FDD-DL band, so it will be difficult to adopt the in-band FMA approach. Additionally, taking into account the system information transmitted in the different subframes, only two could be employed by TDD-UL under the in-band TMA approach.

#### 2.1.2 Efficient use of the CC-based licensed spectrum

LTE allows reconfiguring bands thanks to the CC aggregation concept [10]. Here, it is assumed that the licensed spectrum is divided in CCs, and there is an entity responsible for the dynamic long/medium-term resource allocation that selects the number of required CCs by the MeNB as a function of the traffic demand. In such scenario, the unused CCs can be employed by the TDD SeNB.

FDD MeNB and the TDD SeNB are working at different operational bandwidths, isolated thanks to the respective transmit filters, but ACI between both systems should be evaluated. When reusing the FDD-UL spectrum, possibly the most important interference is the one received at the FDD MeNB and generated by the TDD SeNB transmitting DL signals. The difference on the power levels transmitted by UEs and SeNBs, along with the probability of line-of-sight (LOS), implies that ACI might be significant. This can be mitigated by deploying SeNBs at a lower height than the MeNB, thus reducing the probability of LOS. When reusing the FDD-DL spectrum, the most significant interference is at the TDD SeNB in UL mode, which is generated by FDD MeNB transmitted DL signals.

### 2.2 Resource provisioning

The use of the flexible duplexing concept requires a tight estimation of the amount of spectrum that will not be employed by the FDD MeNB system. A suitable metric that contains this information is the resource utilization (RU). This metric reports the ratio between the total number of resources used by data traffic over the total number of resources available for data traffic. More explicitly, the RU of the *k*-th cell in the *d*-th transmit direction (DL or UL) can be estimated as [11],

$$\rho_k^d = \frac{\text{average offered traffic}}{\text{traffic served}} = \frac{\lambda_k^d L_k^d}{x_k^d} \frac{1}{C_k^d} \tag{1}$$

where $\lambda_k^d$ is the mean packet arrival rate (in packets/s), $L_k^d$ denotes the mean packet length (in bits/packet), $x_k^d$ refers to the number resources, and $C_k^d$ is the average spectral efficiency of the $k$-th cell in the $d$-th transmit direction (in bits/s/resource). Furthermore, the average number of bits in the queue of the $k$-th cell can be modeled as a function of the RU when packet arrival instants follow a Poisson process. In such a case, for $\rho_k^d \leq 1$, we have the following expression [11],

$$W_k^d = \frac{\rho_k^d}{1-\rho_k^d}\frac{l_k^d}{2L_k^d} \qquad (2)$$

where $l_k^d$ denotes the mean of the squared packet length. Thus, a larger RU factor implies a larger average queue size. We can derive the maximum RU for a given target $W_k^d$, and therefore, we can elucidate the required number of resources ($x_k^d$) for a given traffic demand.

## 2.3 Performance evaluation

In this section we present some simulation results for the different access methods presented in Section 2.1. The scenario consists of one FDD MeNB and one TDD SeNB reusing the unused FDD spectrum. The complete scenario and simulation assumptions are included in Table 1. The following techniques are evaluated in a scenario where the MeNB is separated from SeNB 100 meters and different traffic asymmetries are adopted:

- *FMA* (in-band FMA and CC-based FMA): FMA with frequency multiplexing between FDD MeNB and TDD SeNB, including ACI.
- *TMA* (in-band TMA): TMA with time multiplexing between MeNB and SeNB, including 1 guard subframe.
- *only one MeNB,* operating in the paired spectrum.

As performance indicators two metrics are considered: a) the RU presented in (1), measured as the average number of resource blocks needed for the communication over the total number of available resource blocks, and b) the mean of the user throughput (UT) in Mbits/s. The obtained results are presented in Table 2, where the following conclusions can be drawn:

| General system parameter | |
|---|---|
| MeNBs deployment | One sector of a macrocell area with 1 FDD MeNB. Hexagonal deployment of FDD MeNBs. Inter-site distance of 500 m. Interfering MeNBs operate in FDD with normal usage. |
| SeNB deployment | 1 TDD SeNB at a distance 100 m from MeNB |
| UEs deployment | 50 UEs uniformly distributed within the macrocell area |
| Spectrum | Licensed paired FDD: 10 MHz for DL and 10 MHz for UL (i.e. 50 resource blocks (RBs) on each paired band) |
| Frequency carrier | 2.5 GHz |
| Transmit power | 46 dBm (MeNB), 24 dBm (SeNB), 23 dBm (UE) |
| MeNB antenna system | 17 dBi, 3D, Sectorized, 2 antennas |
| SeNB antenna system | 5 dBi, 2D, Omnidirectional, 2 antennas |
| UE antenna system | 0 dBi, 2D, Omnidirectional, 2 antennas |
| Noise figure | 5 dB at MeNB and SeNB, 9 dB at UEs |
| Noise spectral density | -174 dBm/Hz |
| Propagation conditions | Pathloss and shadowing as in [5]. Frequency selective fading follows the typical urban model. |
| **Cell association** | |
| Criterion | UEs are associated to the MeNB or SeNB using the reference signal received power (RSRP) combined with a cell range expansion (CRE) bias. For a fair comparison, the CRE is adjusted so as to get approximately the same number of SUEs (around 10 UEs) for any position of SeNB. |
| **Traffic generation** | |
| Traffic model | FTP model 3. Packets for the same UE arrive according to a Poisson process with arrival rate $\lambda_{DL}$ (DL) and $\lambda_{UL}$ (UL) (in packets/s). Packet size 2 Mbits. |
| Traffic asymmetries | Two traffic asymmetries are used:<br>1) $\lambda_{UL} = 0.1 \times \lambda_{DL}$ : Unused resources in the FDD-UL band.<br>2) $\lambda_{UL} = 10 \times \lambda_{DL}$ : Unused resources in FDD-DL band. |
| **Resource allocation for orthogonal access** | |
| FMA | *FDD-UL band reuse*: 10MHz are distributed among the FDD MeNB UL (30%) and the TDD SeNB (70%). TDD duplexing 7:3, i.e. 7 DL and 3 UL subframes.<br>*FDD-DL band reuse*: 10MHz are distributed among the FDD MeNB DL (14%) and the TDD SeNB (87%). TDD duplexing 1:9, i.e. 1 DL and 9 UL subframes. |
| TMA | TMA is only evaluated for reuse of the licensed FDD-UL band. 10MHz are used by the FDD MeNB UL (30%) and the TDD SeNB (60%) in different time instances. TDD duplexing 4:2, 4 subframes for DL and 2 subframes for UL. |

**Table 1. Simulation assumptions.**

| $\lambda_{UL} = 0.1 \times \lambda_{DL}$ **reuse FDD-UL band** | | RU DL | | RU UL | | UT DL (Mbits/s) | | UT UL (Mbits/s) | |
|---|---|---|---|---|---|---|---|---|---|
| | | *MeNB* | *SeNB* | *MeNB* | *SeNB* | *MeNB* | *SeNB* | *MeNB* | *SeNB* |
| $\lambda_{DL}$=1 | only MeNB | 0.97 | | 0.22 | | 1.84 | | 0.31 | |
| | FMA | **0.86** | 0.22 | 0.15 | 0.07 | **1.97** | **1.97** | 0.31 | 0.30 |
| | TMA | **0.86** | 0.20 | 0.16 | 0.07 | **1.97** | **1.97** | 0.31 | 0.30 |
| $\lambda_{DL}$=1.5 | only MeNB | 0.99 | | 0.33 | | 1.94 | | 0.38 | |
| | FMA | 0.97 | 0.37 | 0.25 | 0.12 | **2.32** | **2.77** | **0.36** | **0.35** |
| | TMA | 0.97 | 0.28 | 0.21 | 0.10 | **2.32** | **2.89** | 0.38 | 0.37 |
| $\lambda_{UL} = 10 \times \lambda_{DL}$ **reuse FDD-DL band** | | RU DL | | RU UL | | UT DL (Mbits/s) | | UT UL (Mbits/s) | |
| | | *MeNB* | *SeNB* | *MeNB* | *SeNB* | *MeNB* | *SeNB* | *MeNB* | *SeNB* |
| $\lambda_{DL}$=0.1 | only MeNB | 0.10 | | 0.98 | | 0.31 | | 1.75 | |
| | FMA | 0.08 | 0.05 | 0.96 | 0.74 | 0.31 | **0.27** | **1.93** | **1.43** |
| $\lambda_{DL}$=0.15 | only MeNB | 0.16 | | 0.99 | | 0.39 | | 1.98 | |
| | FMA | 0.11 | 0.07 | 0.98 | 0.76 | 0.37 | **0.18** | **2.21** | **1.43** |

**Table 2. Simulation results in terms of RU and UT (in Mbits/s) for FDD-UL band reuse and FDD-DL band reuse with *d*=100m.**

When there are unused resources in the FDD-UL band and the ones in the FDD-DL are nearly saturated:

- An improvement of 'UT DL' at both the MeNB and the SeNB is obtained as compared to the case of having only the MeNB active (in green in Table 2-top), as more DL traffic can be served.
- For low traffic loads ($\lambda_{DL}$=1packets/s), thanks to the reuse of the licensed bandwidth for UL (see FMA and TMA), the MeNB DL is not saturated as compared to the case of having only the MeNB active, see 'RU DL MeNB' in Table 2-top.
- The ACI (SeNB to MeNB, SUE to MeNB) in the FMA imposes a lower transmission rate, so that more resources are needed (see 'RU DL SeNB', 'RU UL MeNB', and 'RU UL SeNB'). The UT becomes degraded at high traffic loads due to the activity of SeNB. It generates interference towards the MeNB, so the transmission rate of MUEs is lowered, but being active more time, and impacting negatively to the SeNB when it is in UL (in red in the Table 2-top).This effect is avoided using TMA.

On the other hand, when the unused resources are in the FDD-DL band, results depend on two main impairments: a) *DL interference from neighboring MeNBs* (co-channel external interference) can significantly degrade the system performance, as the MeNB interferes the SeNB transmissions both in DL and UL, b) *ACI at SeNB is important* because transmitters in the neighboring band are MeNBs. From Table 2-bottom we infer that:

- An improvement of the 'UT UL MeNB' is obtained as compared to the case of having only the MeNB active (coloured in green in the Table 2-bottom), as more UL traffic can be served.
- For all traffic loads, the 'UT UL SeNB' is significantly degraded. This is due to the co-channel external interference from neighbouring MeNBs, which significantly impacts on SeNB UL (coloured in red in the Table 2-bottom). In addition, the effect of co-channel external interference also degrades 'UT DL SeNB'.

- There is nearly no impact on the 'UT DL MeNB' and similar values are obtained as compared to the case of having only the MeNB, since all the traffic that arrives to the system is being served.

# 3 Current limitations

In spite of the promising benefits shown by the flexible duplexing concept in Section 2.3, its implementation in the short term must face several challenges.

## 3.1 Regulation

Radio spectrum regulators define which type of transmissions are allowed on different parts of the spectrum. In general, the FDD-UL spectrum can be employed by mobile stations or end-users, but not by base stations. In this regard, a survey about the flexible use of the spectrum was carried out over different regulators [12], concluding that at least in US the flexible duplex concept is allowed in the band 1719-1755 MHz. On the other hand, in Europe (ECC PT1) or Japan (ARIB), the flexible use of UL and DL for FDD bands is not allowed. Nevertheless, the use of SeNB with a transmit power equivalent to the maximum allowed in the UL by regulation would satisfy all technical requirements. In this regard, ECC PT1 is open to receive new results about the benefits of the flexible use of the band.

## 3.2 LTE standard

The following aspects limit the derivation of LTE standard-compliant procedures for the implementation the flexible duplexing concept.

*Operating band definition*. LTE defines a set of operating bands along with its use: FDD (1-32) or TDD (33-44), see [10]. From the comparison of UL-DL FDD bands and the TDD bands (see Table 3), we can observe that with the current definition, not all FDD bands could be reused by TDD systems. It is interesting to notice that those FDD systems operating in band 7 (2500-2570, 2620-2690 MHz available in Europe and Hong-Kong), might adopt the flexible duplex concept by deploying TDD SeNBs operating in band 41. The UEs in the area should just measure the control channels of MeNB and SeNB and decide their association to one of the nodes.

| E-UTRA Operating Band | Uplink (UL) operating band | | Downlink (DL) operating band | |
|---|---|---|---|---|
| | *BS receive, UE transmit* | *Spectrum used as TDD in band* | *BS transmit, UE receive* | *Spectrum used as TDD in band* |
| 1 | 1920 MHz<br>1980 MHz | 36 | 2110 MHz<br>2170 MHz | × |
| 2 | 1850 MHz<br>1910 MHz | 33, 35 | 1930 MHz<br>1990 MHz | 36 |
| 3 | 1710 MHz<br>1785 MHz | × | 1805 MHz<br>1880 MHz | 35, 39 |
| 4 | 1710 MHz<br>1755 MHz | × | 2110 MHz<br>2155 MHz | × |
| 5 | 824 MHz<br>849 MHz | × | 869 MHz<br>894MHz | × |
| 6 | 830 MHz<br>840 MHz | × | 875 MHz<br>885 MHz | × |
| 7 | 2500 MHz<br>2570 MHz | 41 | 2620 MHz<br>2690 MHz | 41 |
| 8 | 880 MHz<br>915 MHz | × | 925 MHz<br>960 MHz | × |
| 9 | 1749.9MHz<br>1784.9 MHz | × | 1844.9MHz<br>1879.9 MHz | 35, 39 |
| 10 | 1710 MHz<br>1770 MHz | × | 2110 MHz<br>2170 MHz | × |
| 11 | 1427.9MHz<br>1447.9 MHz | × | 1475.9MHz<br>1495.9 MHz | 32 |
| 12 | 699 MHz<br>716 MHz | 44 | 729 MHz<br>746 MHz | 44 |
| 13 | 777 MHz<br>787 MHz | 44 | 746 MHz<br>756 MHz | 44 |
| 14 | 788 MHz<br>798 MHz | 44 | 758 MHz<br>768 MHz | 44 |
| 17 | 704 MHz<br>716 MHz | 44 | 734 MHz<br>746 MHz | 44 |
| 18 | 815 MHz<br>830 MHz | × | 860 MHz<br>875 MHz | × |
| 19 | 830 MHz<br>845 MHz | × | 875 MHz<br>890 MHz | × |
| 20 | 832 MHz<br>862 MHz | × | 791 MHz<br>821 MHz | 44 |
| 21 | 1447.9MHz<br>1462.9 MHz | 32 | 1495.9 Hz<br>1510.9 MHz | × |
| 22 | 3410 MHz<br>3490 MHz | 42 | 3510 MHz<br>3590 MHz | 42 |
| 23 | 2000 MHz<br>2020 MHz | 34 | 2180 MHz<br>2200 MHz | × |
| 24 | 1626.5MHz<br>1660.5 MHz | × | 1525 MHz<br>1559 MHz | × |
| 25 | 1850 MHz<br>1915 MHz | 39 | 1930 MHz<br>1995 MHz | 36 |
| 26 | 814 MHz<br>849 MHz | × | 859 MHz<br>894 MHz | × |
| 27 | 807 MHz<br>824 MHz | × | 852 MHz<br>869 MHz | × |
| 28 | 703 MHz<br>748 MHz | 44 | 758 MHz<br>803 MHz | 44 |
| 29 | N/A | × | 717 MHz<br>728 MHz | 44 |
| 30 | 2305 MHz<br>2315 MHz | 40 | 2350 MHz<br>2360 MHz | 40 |
| 31 | 452.5 MHz<br>457.5 MHz | × | 462.5 MHz<br>467.5 MHz | × |
| 32 | N/A | × | 1452 MHz<br>1496 MHz | × |

**Table 3. Candidate E-UTRA FDD operating bands that might be used as TDD.**

*PUCCH in the FDD UL band*. Current design of LTE FDD systems places the PUCCH in the resource blocks located at the edge of the band [13]. Those resources are devoted to transmit system information for UE. This system constraint limits the flexibility for seizing the unused time-frequency resources in-band TMA approach described in Section 2.1.1.

*Frame structure in FDD DL band*. Even in situations of low DL traffic, the information necessary to operate the system (to be more specific, synchronization signals and system information and paging) needs to be transmitted by the FDD-DL cell so that terminals can find and connect to a cell. In FDD, the subframes where such information is provided are subframes 0, 4, 5, and 9 within an LTE frame composed of 10 subframes. Therefore, these subframes must be used by the FDD cell and cannot be used by a TDD cell. Similarly, the TDD SeNB needs also to transmit the information necessary to operate the system. In TDD, such information is provided through subframes 0, 1 (special subframe), 5, and 6 (special subframe) within a frame of 10 subframes. It turns out that only 2 subframes could be reused for data transmission in the in-band TMA approach. On the other hand, the SeNB should be deployed in a narrower bandwidth placed at one side of the band for the in-band FMA approach because all synchronization signals of FDD DL occupy the central part of the band.

*Carrier Aggregation.* Currently, the 3GPP standard does not allow tackling situations where the traffic asymmetry is higher in the FDD-UL band because, by definition, the CCs in the UL must be smaller than in the DL [13]. This feature does not allow extending the concept explained in Section 2.1.2 to re-use the underutilized FDD spectrum in the FDD-DL band.

# 4 Technical challenges

There are important technical challenges that require further investigation for the deployment of the flexible duplex concept:

*Timing offset adjustments*. In the heterogeneous scenario described in [14] consisting of one MeNB and one SeNB sharing the same band and the same duplexing, it was shown that UEs have to advance their UL transmissions (*pre-compensation*) not only by taking into account the propagation delay with the SeNB, but also by considering the propagation delay with the MeNB and the receive frame boundary. The cyclic prefix in OFDM systems allows combatting this issue, in addition to maintain the orthogonality among subcarriers. However, in the flexible duplexing concept, SeNBs work in TDD while MeNB is FDD-UL, which means that synchronization is more challenging because the SeNB DL transmission should be pre-compensated by taking into account the neighboring FDD MeNBs and TDD SeNB working in UL.

*Traffic-aware resource management*. The RU presented in Section 2.2, in addition to allow the identification of the required spectrum, is a useful metric to derive resource provisioning schemes in a multi-cell scenario, (e.g. multiple TDD SeNBs that exploit the spectrum released by the FDD MeNB). An efficient resource provisioning is such that the resources are distributed among all the cells in a balanced way and try to avoid very different occupancies. In this sense, a suitable optimization criterion is the minimization of the maximum RU among cells, such that resources are fairly distributed and more resources are given to those cells with larger traffic loads and/or those cells experiencing greater delays. For example, long-term graph colouring-based resource provisioning schemes are presented in [15] with the objective of optimizing the RU factors of multiple TDD SeNBs when either orthogonal access is assumed or reuse access among non-interfering SeNBs is considered.

*Interference cancellation in opportunistic multiple access*. If signal time offsets are pre-compensated to avoid the time misalignments with FDD MeNB, then the opportunistic multiple access technique can be used by the TDD SeNBs, see Section 2.1.1. The interference received by the MeNB from the transmitting SeNBs might be very high due to LOS in the MeNB-SeNB link. A successive interference canceller would alleviate the effect of interference, but this will depend on the distance between MeNB-SeNB and the bit rate selected by the SeNB (DL transmissions to a SUE) and MUE (UL transmissions to MeNB).

# 5 Conclusions

The flexible duplexing concept allows improving the system efficiency of paired-based systems by using TDD SeNBs. When allowing a TDD SeNB to operate in the underutilized FDD-UL band, we have observed that the DL user throughput is improved, reducing also the congestion of the MeNB. The effect of the ACI for the FMA schemes becomes relevant when: a) the SeNB is close to the MeNB, b) SeNB transmits with high power, and c) the activity of UEs and SeNB is high. On the other hand, when a TDD SeNB reuses the underutilized FDD-DL band the potential gains are reduced because of the external interference coming from neighboring MeNB that are transmitting DL signals. The benefits of the flexible duplexing concept can be enlarged when combined with the deployment of multiple SeNBs and the application of interference management techniques, but its actual implementability is also tied to the limitations imposed by the regulation and standards.

# 6 Acknowledgments

This work has been supported by Huawei Technologies Co., Ltd.

# 7 References

[1] DIGITALEUROPE: Call for Timely Harmonisation of the 1452-1492MHz and 2300-2400MHz Bands to Support Delivery of the EU Radio Spectrum Policy Programme Objectives, Brussels, 21st February 2012

[2] Qualcomm: Wireless broadband future and challenges, Oct 27th, 2010

[3] 3GPP RP-140062, "Motivation of New SI proposal: Evolving LTE with Flexible Duplex for Traffic Adaptation", March 2014

[4] W. Lei, Z. Mingyu, W. Rongui, "Evolving LTE with Flexible Duplex", in IEEE Proceedings Globecom 2013

[5] 3GPP TR 36.828, Further enhancements to LTE Time Division Duplex for Downlink-Uplink interference management and traffic adaptation. Release 11. June 2012

[6], ECC Report 119. Coexistence between Mobile Systems in the frequency band at the FDD/TDD Boundary. June 2008

[7] Real Wireless. Low-power shared access to spectrum for mobile broadband. Ofcom project MC/073. March 2011

[8] R. Berangi, S. Saleem, M. Faulkner, W. Ahmed, “TDD cognitive radio femtocell network (CRFN)”, in Proc. IEEE Int. Symposium of Personal, Indoor and Mobile Radio Communications, Sep. 2011

[9] 3GPP. R1-134295, “FDD-TDD CA/Dual Connectivity solution exploiting traffic asymmetry in duplex-neutral bands” (Oct-2013).

[10] 3GPP TS 36.101, “User Equipment radio transmission and reception”, Release 13, July 2015

[11] Kumar, D. Manjunath, and J. Kuri, Communication Networking: an analytical approach. Morgan Kaufmann Publishers, Elsevier, 2004

[12] 3GPP TR 36.882, “Study on regulatory aspects for flexible duplex for E-UTRAN”, Release 13, June 2015

[13]. E. Dahlman, S. Parkvall, J. Sköld, *4G LTE/LTE-Advanced for Mobile Broadband*, Academic Press, 2011

[14] N. Himayat, et al., “Synchronization Uplink Transmissions from Femto AMS”, IEEE C802.16m-09/30775r2, January 2010

[15] S. Lagen, O. Muñoz, A. Pascual-Iserte, J. Vidal, A. Agustin, "Long-term provisioning of radio resources based on their utilization in dense OFDMA networks", IEEE Int. Symp. on Personal, Indoor and Mobile Radio Communications., Valencia (Spain), Sep. 2016.